\documentclass[iop,apjl,12pt]{emulateapj}
\usepackage{times}
\usepackage[normalem]{ulem}
\usepackage{epsfig}
\usepackage{natbib}
\usepackage{amsfonts}
\usepackage{amsmath}
\usepackage{multirow}
\usepackage{enumerate}
\usepackage[usenames]{color}
\usepackage[plainpages=false, colorlinks=true, anchorcolor=blue, linkcolor=blue, citecolor=blue, bookmarks=false]{hyperref}
\providecommand{\abs}[1]{\lvert#1\rvert}
\newcommand{\fheat}{$f_\textrm{heat}$}

\def\sun{_\odot}

\def\Dwa{$\,$\uppercase\expandafter{\romannumeral5}$\,$}

\def\sless{\lower2pt\hbox{$\buildrel {\scriptstyle <}
   \over {\scriptstyle\sim}$}}

\def\sgreat{\lower2pt\hbox{$\buildrel {\scriptstyle >}
   \over {\scriptstyle\sim}$}}
\def\sharpnull#1{}

\newcommand{\subdate}{2013 September 10}
\newcommand{\shortauth}{Couch \& Ott}
\newcommand{\slugcom}{Submitted to ApJ Letters on \subdate}
\slugcomment{\slugcom}

\lefthead{\sc \footnotesize \slugcom \hfill \shortauth}
\righthead{\sc \footnotesize \slugcom \hfill \shortauth}

\begin{document}

\title{Revival of The Stalled Core-Collapse Supernova Shock \\
Triggered by Precollapse Asphericity in the Progenitor Star}

\author{Sean M. Couch\altaffilmark{1,*}}
\author{Christian D. Ott\altaffilmark{2,3,+}}
  \altaffiltext{1}{Flash Center for Computational Science, Department of Astronomy \& Astrophysics, University of Chicago, Chicago, IL, 60637, smc@flash.uchichago.edu}
  \altaffiltext{2}{TAPIR, Mailcode 350-17,
  California Institute of Technology, Pasadena, CA 91125, USA, 
  cott@tapir.caltech.edu}
\altaffiltext{3}{Kavli Institute for the Physics and
 Mathematics of the Universe (Kavli IPMU WPI), The University of Tokyo, Kashiwa, Japan}
\altaffiltext{*}{Hubble Fellow}
\altaffiltext{+}{Alfred P. Sloan Research Fellow}

\begin{abstract}
Multi-dimensional simulations of advanced nuclear burning stages of
massive stars suggest that the Si/O layers of presupernova stars
harbor large deviations from the spherical symmetry typically assumed
for presupernova stellar structure. We carry out three-dimensional
core-collapse supernova simulations with and without aspherical
velocity perturbations to assess their potential impact on the
supernova hydrodynamics in the stalled shock phase.  Our results show
that realistic perturbations can qualitatively alter the postbounce
evolution, triggering an explosion in a model that fails to explode
without them. This finding underlines the need for a multi-dimensional
treatment of the presupernova stage of stellar evolution.
\end{abstract}


\section{Introduction}
The core-collapse supernova (CCSN) phenomenon is fundamentally
multi-dimensional.  Axisymmetric (2D) and three-dimensional (3D)
simulations have shown that convection and the standing accretion
shock instability (SASI) robustly break spherical symmetry in the
pre-explosion stalled-shock phase (see, e.g.,
\citealt{Couch:2013leak,Couch:2013fh,ott:13a,Dolence:2013iw,Hanke:2013kf,Takiwaki:2013ui}
for recent 3D simulations).  The propagation of artificially initiated
explosions through the progenitor envelope found that symmetry is
broken by Rayleigh-Taylor and Richtmyer-Meshkov instabilities (e.g.,
\citealt{Couch:2009bu,hammer:10,joggerst:10}). The conclusions of
these simulations are backed up by observations of asphericities in
local supernova remnants \citep{vink:12}, by spectropolarimetry of
distant CCSNe \citep[][and references therein]{{Wang:2008bk},
  chornock:11}, and by pulsar kicks
  \citep[e.g.,][]{hobbs:05}.

For initial conditions based on 1D stellar evolutionary models, the 
breaking of spherical symmetry after the initial collapse and
bounce of the inner core is widely appreciated. 
Stars, however, are not truly spherical.  Yet, the
current state-of-the-art in CCSN progenitor evolution is 1D.
Such models resort to various kludges to account for multi-D phenomena such as
convection, rotation, and magnetic fields (see \citealt{langer:12} for
a review). Exploratory explicit multi-D hydrodynamics simulations of
the Si/O-shell burning stage prior to core collapse
\citep{bazan:98,Meakin:2007dj,Arnett:2011ga} have shown that violent
fluctuations about the mean turbulent flow can lead to low-mode
deviations from spherical symmetry.  These fluctuations may also
trigger eruptions that partially unbind the stellar envelope, leading
to precursor transients weeks to months prior to core collapse
(\citealt{smith:13b}, but also see \citealt{quataert:12}). This has now
been observed for multiple CCSNe. The fluctuations and their
consequences cannot be captured by the standard mixing-length approach
for convection and time-implicit stellar evolution codes
\citep{smith:13b}.

The perturbations caused by Si/O shell burning fluctuations are part
of the supersonically collapsing outer core and may be amplified
during collapse \citep{lai:00}. They reach the stalled shock
$\sim$$100-300\,\mathrm{ms}$ after bounce, depending on the structure
of the progenitor. At this time, neutrino-driven convection and/or
SASI are active and may be affected by spatial variations in the
accretion flow.  \cite{bh:96} were the first to carry out 2D collapse
simulations of a progenitor whose density outside $0.9 M_\odot$ was
decreased by $15\%$ within a $20^\circ$ wedge of the pole. They found
an early explosion in the direction of the perturbation and a
hydrodynamically kicked protoneutron star.
\cite{fryer:04kick}, studied similarly large $\ell = 1$ perturbations
applied globally, or only in the Si/O layers, using 3D smooth particle
hydrodynamics. He also found neutron star kicks and explosion
asymmetries, though of smaller magnitude than observed in 2D.

In this Letter, we examine the role of perturbations on the
explosion mechanism {\it itself}. We carry out 3D simulations of the
postbounce evolution of a nonrotating $15$-$M_\odot$ progenitor star.
Unlike previous work, we apply momentum-preserving tangential velocity
perturbations with spatial frequency and magnitude motivated by
\cite{bazan:98} and \cite{Arnett:2011ga}. We also carry out
unperturbed control simulations for comparison. Our results
demonstrate that asphericities in the Si/O layer increase the strength
of turbulence behind the stalled but dynamic shock. This creates
conditions more favorable for shock expansion. We show that
the perturbations can trigger explosion in a model
that would not explode otherwise.

\section{Methods and Setup}

We simulate 3D Newtonian CCSN postbounce evolution using the FLASH
simulation framework \citep{{Fryxell:2000em}, {Dubey:2009wz},
  Lee:2013flash}.\footnote{Available at
  \url{http://flash.uchicago.edu}.}  Our basic numerical approach is
described by \citet{Couch:2013leak} and
\citet{Couch:2013fh}.  We use the multispecies neutrino leakage scheme
of \citet{OConnor:2010bi}, whose 3D version was also employed in
\citet{{Ott:2012ib}, {ott:13a}}.  The neutrino leakage scheme includes
a multiplicative factor, \fheat, in the neutrino heating source term,
which can be adjusted to yield more efficient neutrino heating (i.e.,
\fheat $> 1$).  The leakage scheme with \fheat = 1.00 is tuned to match
the multiangle, multigroup full neutrino transport simulations of
\citet{Ott:2008gn}.
In all simulations reported here, we use 3D Cartesian geometry with a
finest grid spacing $dx_{\rm min} = 0.49$ km.  Using adaptive mesh
refinement, we achieve a pseudo-logarithmic grid by decrementing the
maximum allowed refinement level as a function of radius.  The typical
effective ``angular'' resolution is 0.37$^\circ$.

We use a single progenitor model, the $15$-$M_\odot$ star of
\citet[][]{Woosley:2007bd}.  In order to study the dependence of 3D
CCSN simulations on asphericities extant in the progenitor, we apply
perturbations to the 1D stellar profile.  We seed perturbations that
are convolutions of sinusoidal functions of radius and angle.  For
simplicity, we perturb {\it only} the velocity in the spherical
$\theta$-direction and leave all other variables
untouched.  The form of the sinusoidal perturbation to
$v_\theta$ is
\begin{equation}
  \delta v_\theta = M_{\rm pert} c_S \sin [(n-1)\theta] \sin [(n-1) \zeta] \cos (n \phi)\,,
  \label{eq:perts}
\end{equation}
where $M_{\rm pert}$ is the peak Mach number of the perturbations, $c_S$ is the
local adiabatic sound speed, $n$ is the number of nodes in the
interval $\theta=[0,\pi]$, and $\zeta = \pi (r - r_{\rm
  pert,min})/(r_{\rm pert,max} - r_{\rm pert,min})$.  The
perturbations are only applied within a spherical shell with radial
limits $r_{\rm pert,min} < r < r_{\rm pert,max}$.  We scale the
perturbations with local sound speed so that the peak amplitudes of
the perturbations are constant in {\it Mach} number, not absolute
velocity.  This results in higher-speed perturbations at
smaller radii where the sound speeds are larger.  Importantly, for odd
node numbers, Eq.~(\ref{eq:perts}) results in {\it zero} net
momentum contribution to the inital conditions.  We have verified this
experimentally to  machine-precision.

\vspace*{0.4cm}

\section{Results}

\begin{figure}[tb]
\centering
\includegraphics[width=3.4in, trim= 1.5in 1.25in .75in 1in, clip]{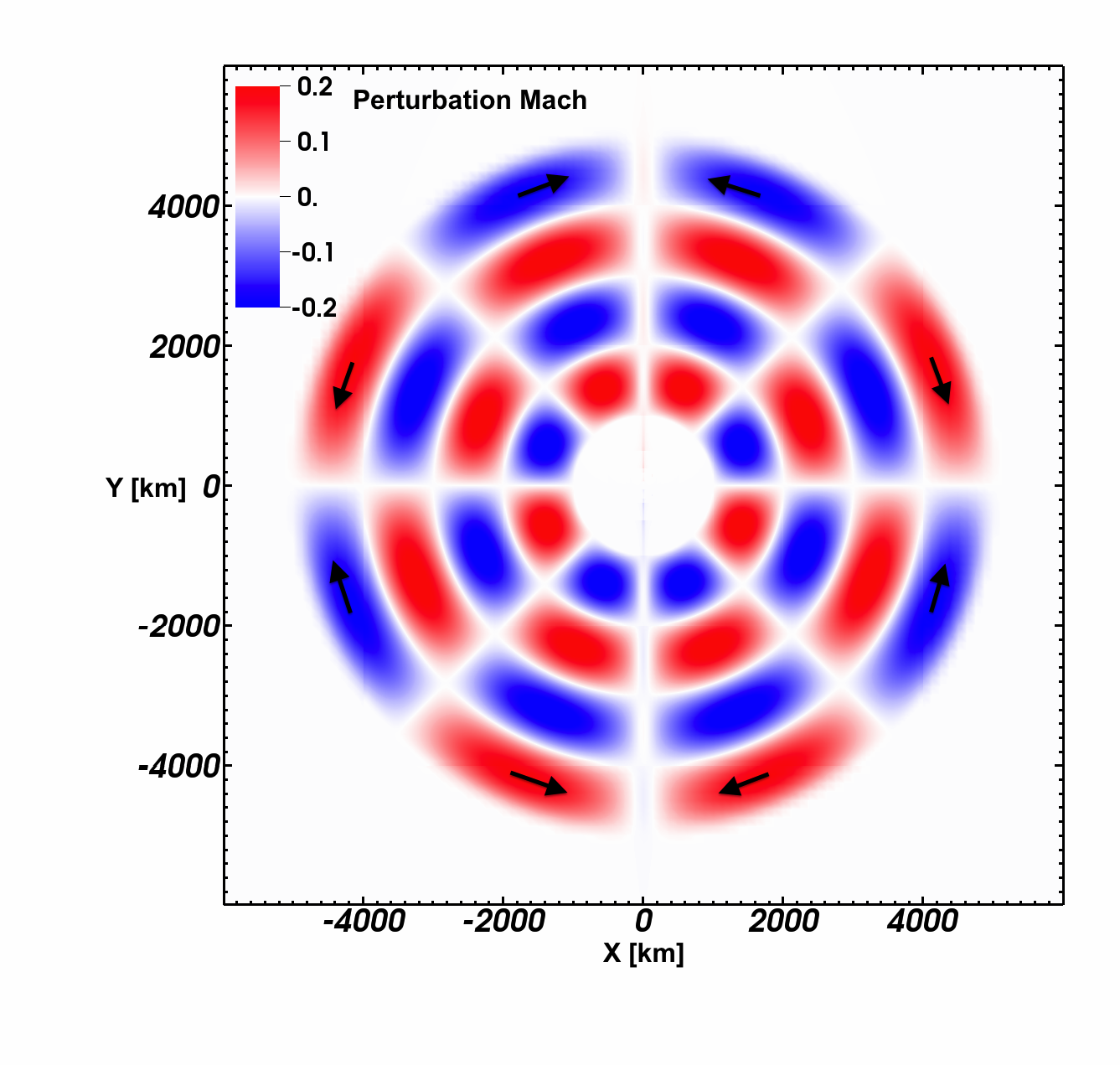}
\caption{ Example of the initial $\theta$-velocity perturbations
  applied in this study.  Shown is the a meridional slice of the Mach
  number of the $\theta$-direction velocity.  The arrows in the outer
  ring of perturbations show the local velocity directions. }
\label{fig:perturbs}
\end{figure}

We start our 3D simulations from the results of 1D simulations at 2 ms
after core bounce, and it is at this point that we apply the
perturbations given by Eq.~(\ref{eq:perts}).  In the results we
discuss here, we use a node count $n=5$ and peak perturbation Mach
number $M_{\rm pert} = 0.2$.  This establishes large-scale
perturbations that are similar in extent and speed to some convective
plumes found in multi-D progenitor burning simulations
\citep{{Meakin:2007dj}, {Arnett:2011ga}}.  We choose $r_{\rm
  pert,min}$ to correspond to the inner edge of the silicon shell
(i.e., the outer edge of the iron core).  For this progenitor at the
time of core bounce, this corresponds to a radius of $\sim1000$ km.
We set $r_{\rm pert,max} = 5000$ km, which is sufficiently large to
never reach the shock during the simulated time period.  Figure
\ref{fig:perturbs} shows a pseudo-color plot of the perturbations used
in this study.

We present the results of four 3D simulations, two perturbed and two
unperturbed.  We use two different heat factors for both perturbed and
unperturbed case: \fheat = 1.00 and a slightly enhanced heating case
with \fheat = 1.02.  We refer to the simulations using the scheme
n[{\it node count}]m[{\it initial perturbation Mach number, times ten}]
\fheat [{\it heat factor}], such that the perturbed model with
enhanced heat factor is referred to as `n5m2 \fheat 1.02.'

\begin{figure}[htb]
\centering
\small\addtolength{\tabcolsep}{-5pt}
\begin{tabular}{cc}
  \includegraphics[width=1.7in,trim= 4.5in 3.in 4in 2.7in, clip]{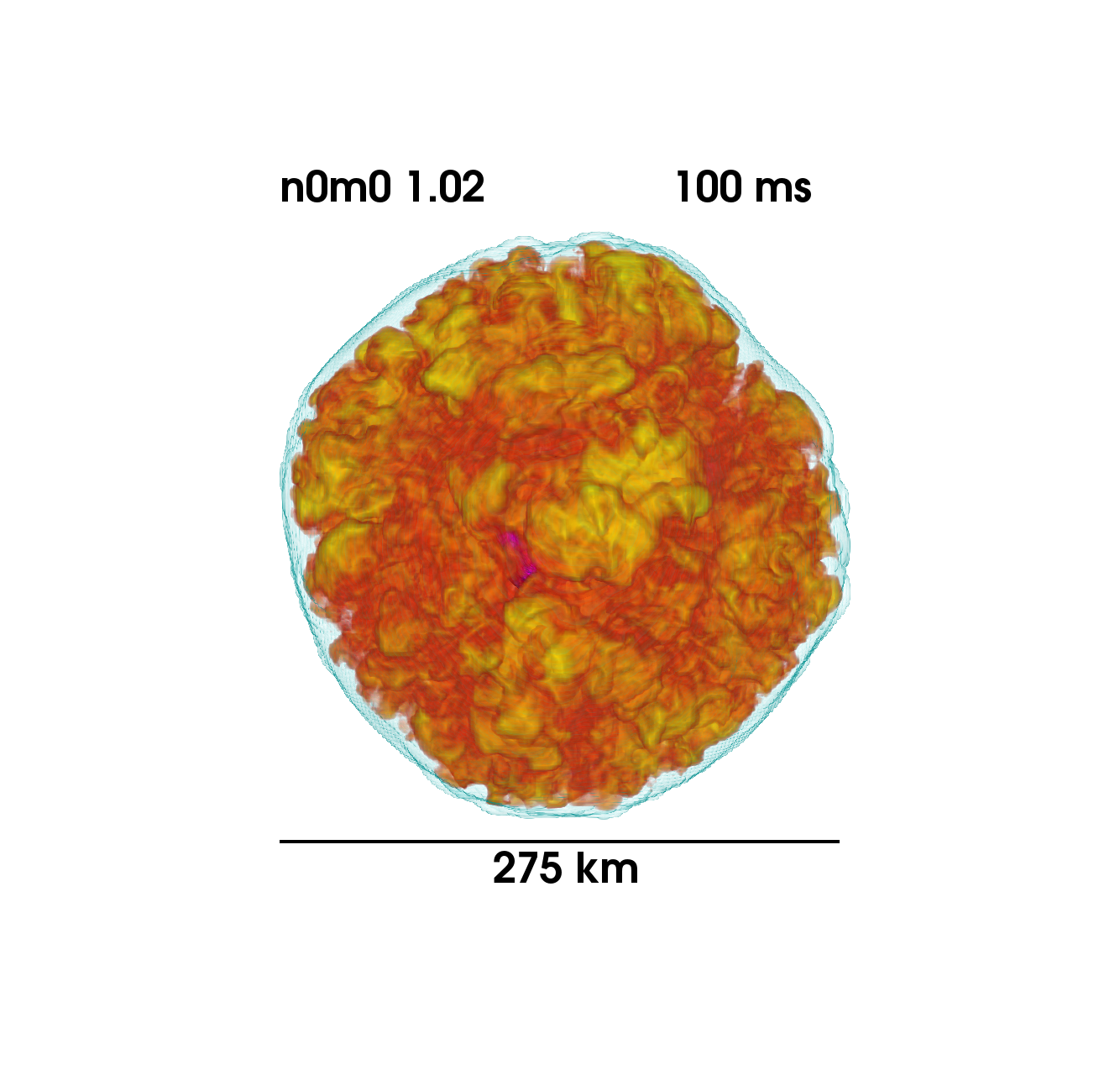}  &
  \includegraphics[width=1.7in,trim= 4.5in 3.in 4in 2.7in, clip]{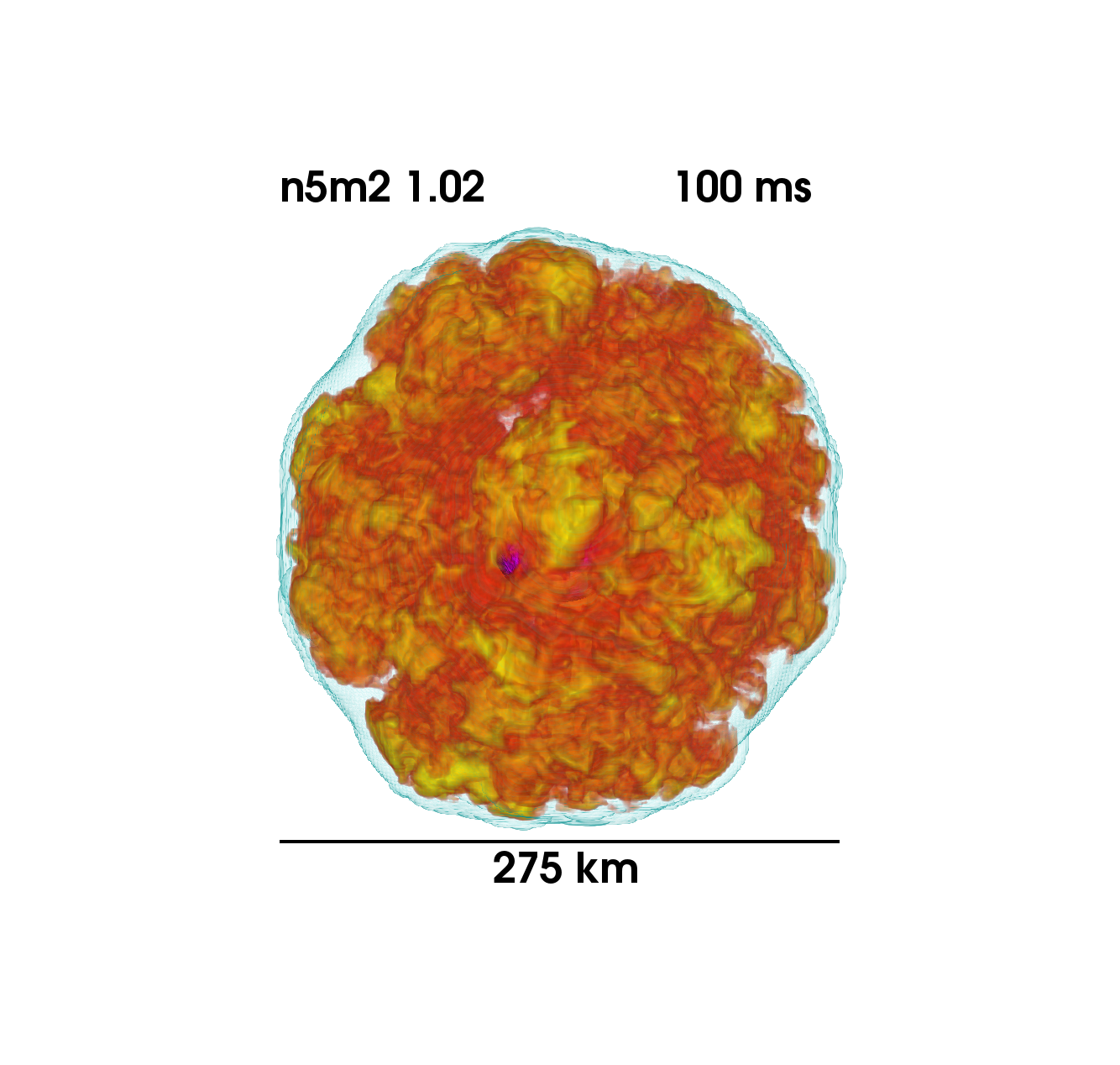}  \\
  \includegraphics[width=1.7in,trim= 4.5in 3.in 4in 2.7in, clip]{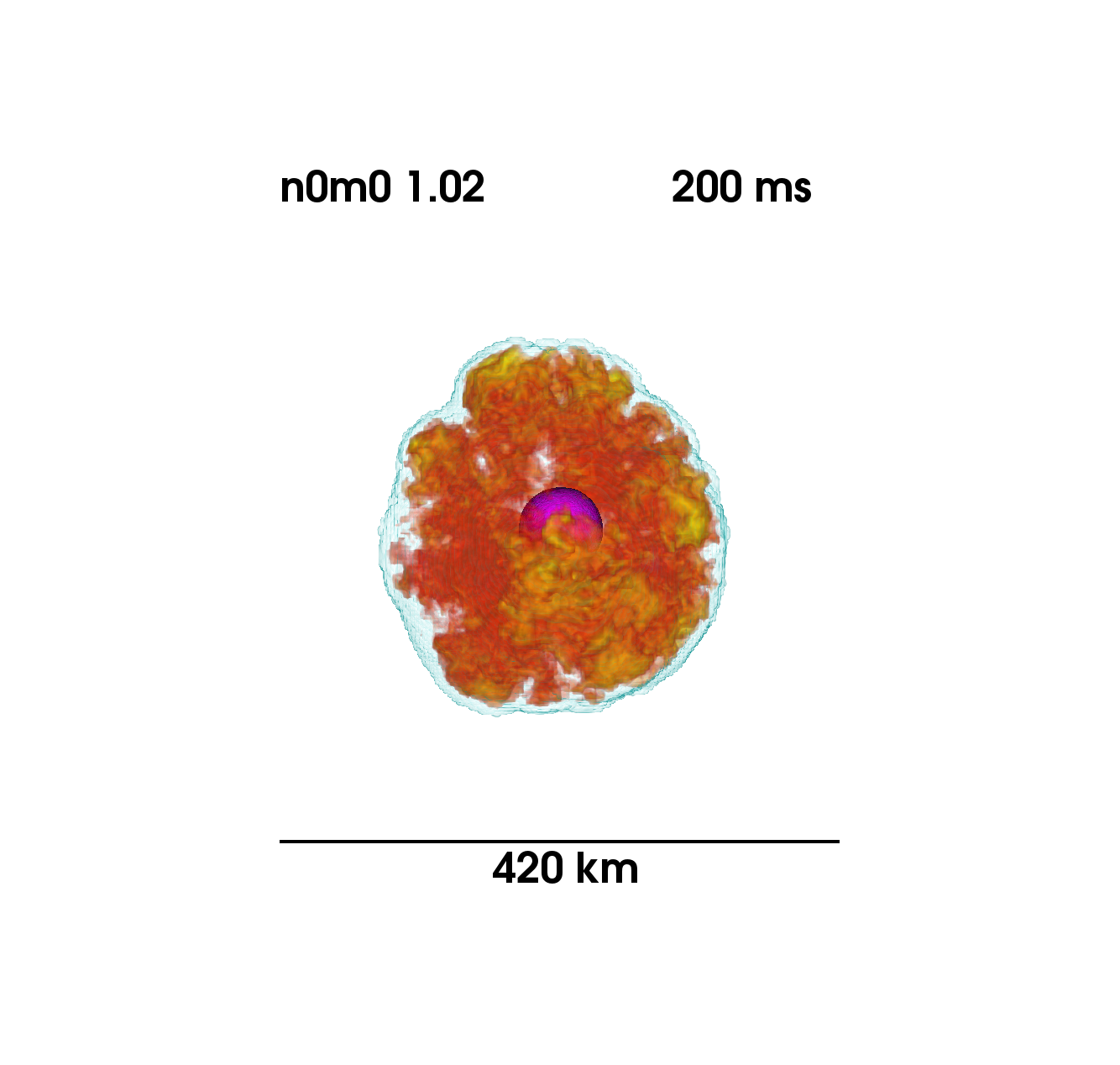}  &
  \includegraphics[width=1.7in,trim= 4.5in 3.in 4in 2.7in, clip]{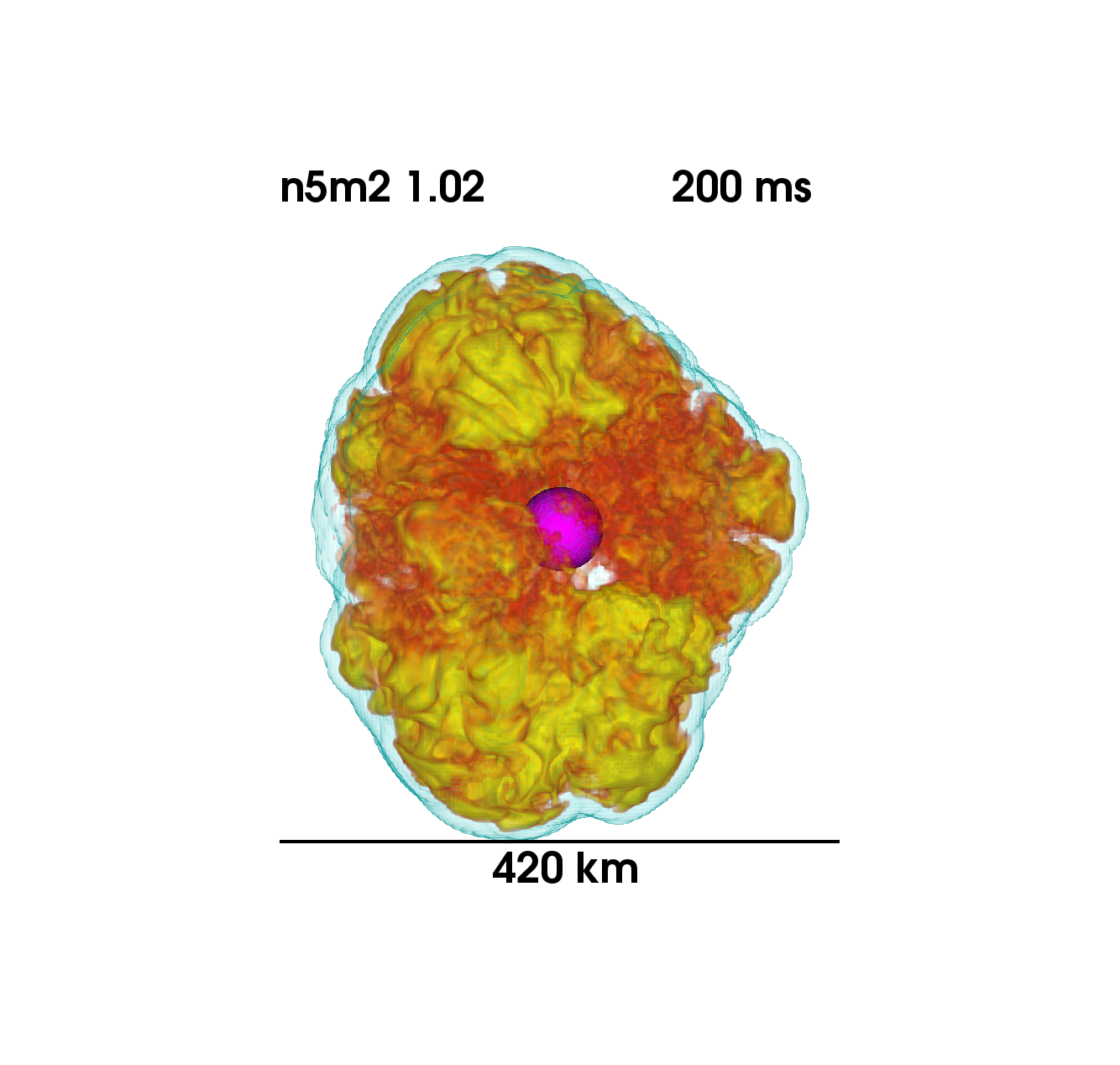}  \\
  \includegraphics[width=1.7in,trim= 4.5in 3.in 4in 2.7in, clip]{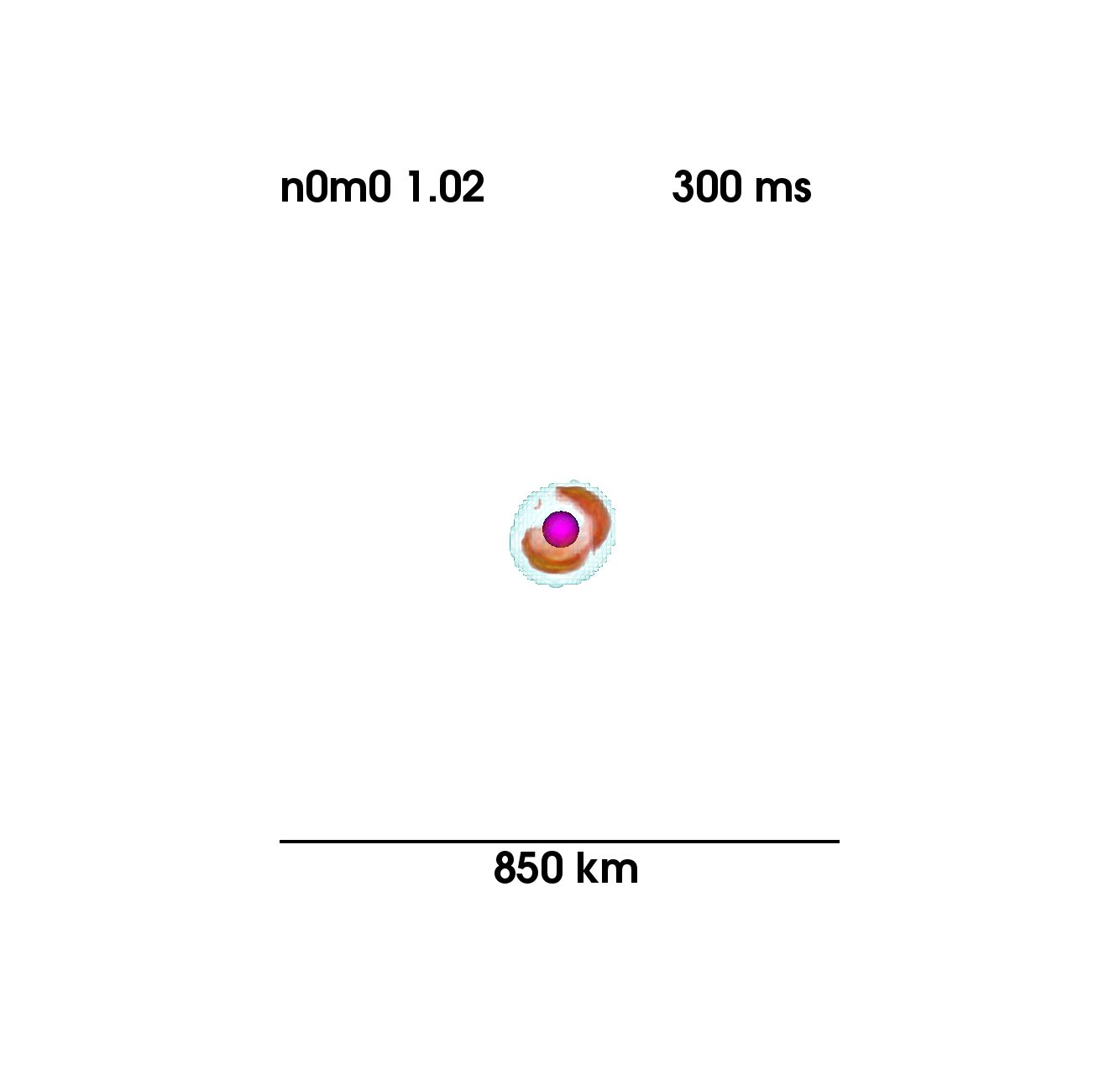}  &
  \includegraphics[width=1.7in,trim= 4.5in 3.in 4in 2.7in, clip]{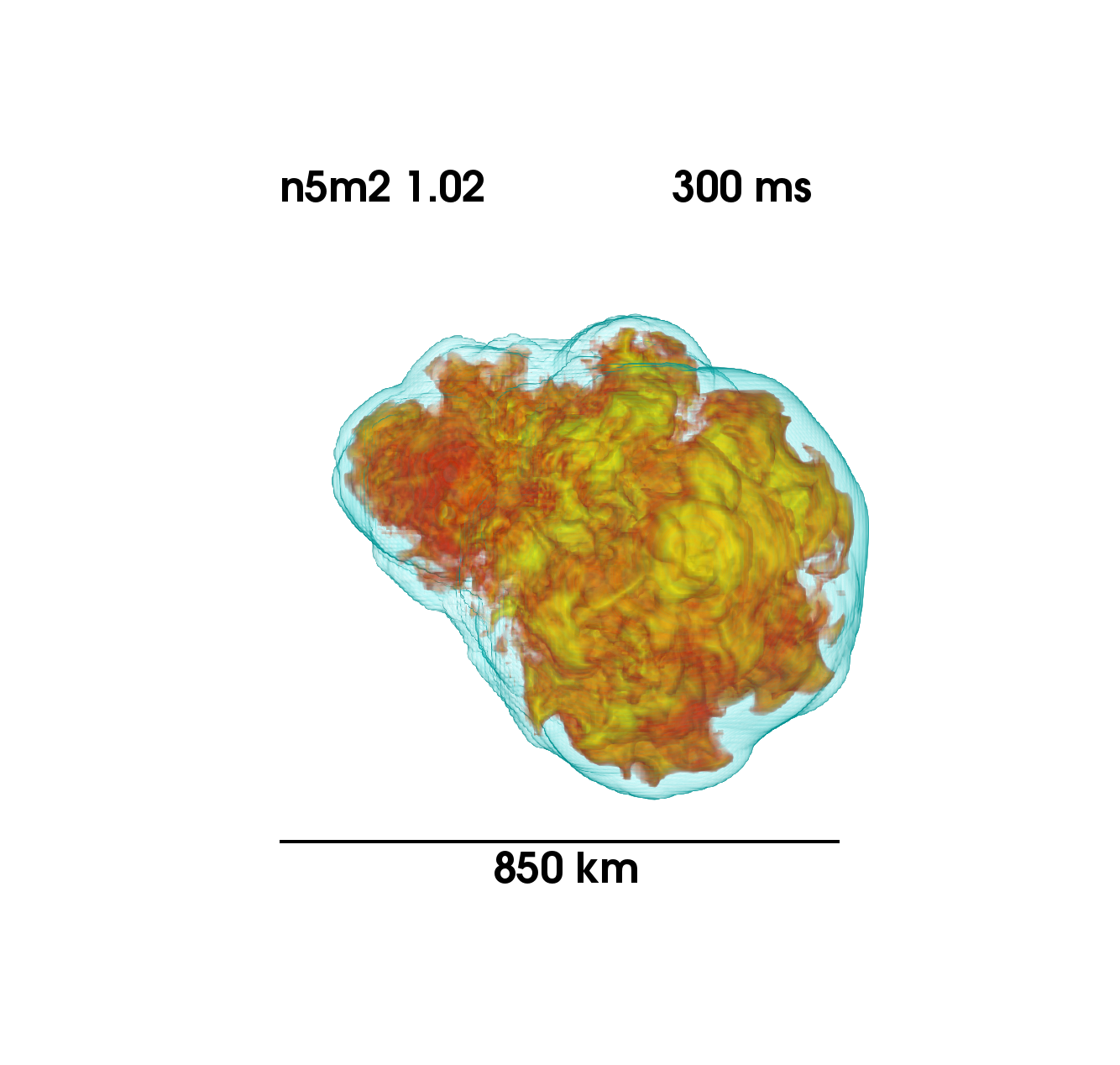} 
\end{tabular}
\caption{ Volume renderings of entropy for models n0m0 \fheat 1.02
  (left column) and n5m2 \fheat 1.02 (right column) at three different
  postbounce times, from top to bottom: 100 ms, 200 ms, and 300 ms.
  The spatial scale is noted at the bottom of each pane and increases
  with time.  The PNS is visible in the center of the renderings,
  marked by a magenta constant-density contour with value $10^{12}$ g
  cm$^{-3}$.
}
\label{fig:volRend}
\end{figure}

We find that introducing plausibly-scaled velocity perturbations in
the Si shell of the progenitor star can trigger a successful explosion
for cases in which an unperturbed simulation fails.  Figure
\ref{fig:volRend} shows several entropy volume renderings for models
n0m0 \fheat 1.02 and n5m2 \fheat 1.02 at three postbounce times.  The
only difference between these two models is the presence of initial
velocity perturbations in the Si/O layer.  Model n5m2 \fheat 1.02
results in continued runaway shock expansion and asymmetric explosion,
as clearly shown, while model n0m0 \fheat 1.02 fails to explode and
the shock recedes to small radii.  At 100 ms, only shortly after the
perturbations have reached the shock, both simulations are quite
similar showing strong convection following the preceding period of
shock expansion.  By 200 ms, however, differences in the models are
obvious.  The shock has already begun to recede in n0m0 \fheat 1.02
while model n5m2 \fheat 1.02 has retained a large shock radius and is
on the verge of runaway shock expansion.  The last frames show the
final states of the two simulations.  Model n5m2 \fheat 1.02 has
exploded, resulting in a large, asymmetric shock structure, while the
shock has fallen back to $\sim$100 km in model n0m0 \fheat 1.02.

\begin{figure}[htb]
\centering
\includegraphics[width=3.4in,trim= 0in 0.15in 0.3in 0.45in,clip]{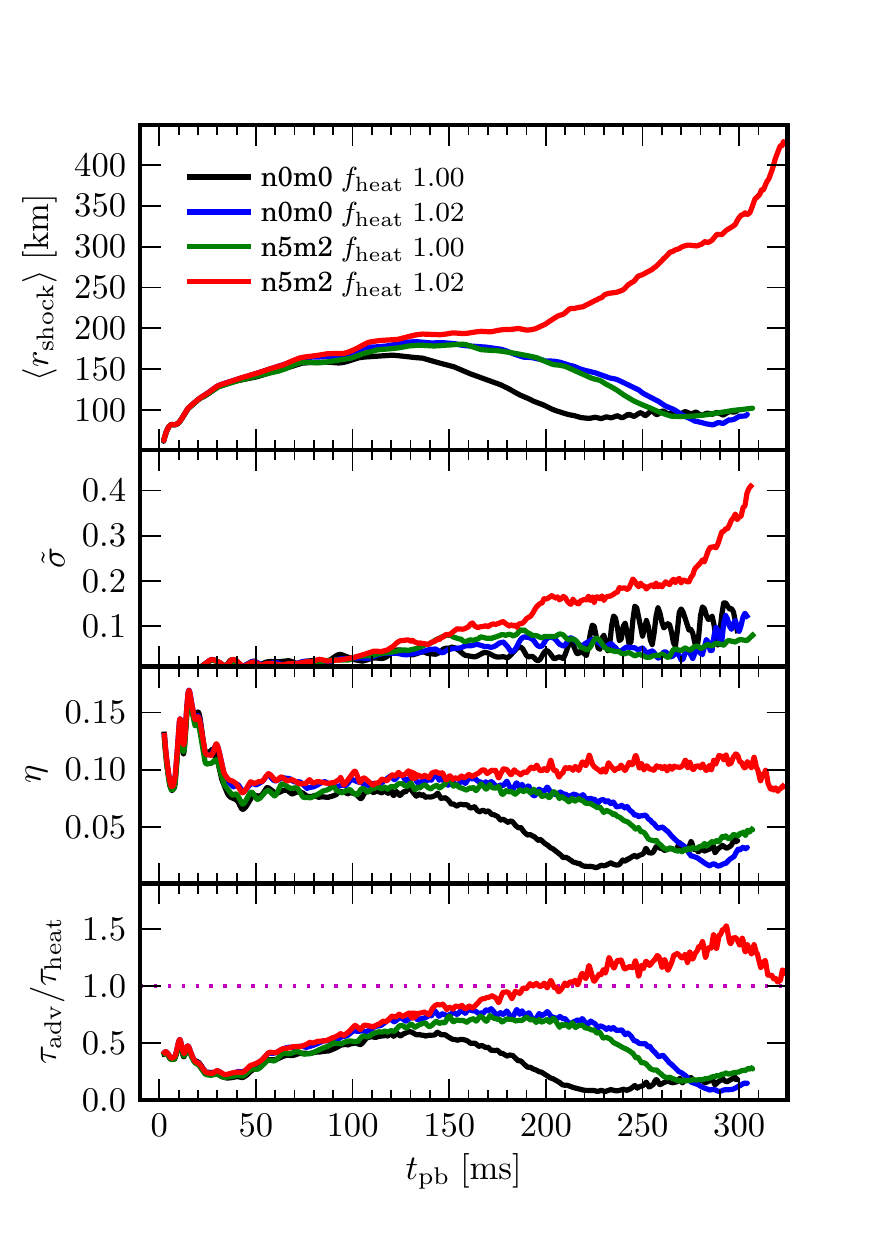} 
\caption{ Time evolution of the global explosion diagnostics for
  our simulations.  Four 3D simulations are shown: unperturbed models
  with \fheat 1.00 (black lines) and 1.02 (blue lines), and perturbed
  models with \fheat 1.00 (green lines) and 1.02 (red lines).  The top
  panel shows the average shock radius.  The second panel shows the
  normalized standard deviation of the shock radius, $\tilde{\sigma} =
  \langle r_{\rm shock} \rangle^{-1} [(4 \pi)^{-1} \int d\Omega
    (r_{\rm shock} - \langle r_{\rm shock} \rangle)^2]^{1/2}$.  The
  third panel shows the heating efficiency, $\eta = Q_{\rm net}
  (L_{\nu_e}+L_{\bar{\nu_e}})^{-1}$.  The bottom panel shows the ratio
  of advection-to-heating time scales.  
}
\label{fig:rshock}
\end{figure}

In Figure \ref{fig:rshock}, we present the time evolutions of several
global metrics for our four 3D simulations.  The top panel of
Fig.~\ref{fig:rshock} shows the average shock radius.  All models,
with the exception of n5m2 \fheat 1.02, fail to explode.  Compared
with the control case, n0m0 \fheat 1.00, both n0m0 \fheat 1.02 and
n5m2 \fheat 1.00 show longer stalled-shock phases prior to shock
recession.  These two intermediate cases, despite employing different
heat factors, show remarkably similar average shock radius histories.
In the case of the successful explosion, n5m2 \fheat 1.02, the average
shock radius remains extremely similar to the comparable unperturbed
model, n0m0 \fheat 1.02, until about 100 ms after bounce.  The average
shock radius of n5m2 \fheat 1.02 remains relatively constant just
below 200 km until $t_{\rm pb}\sim200$ ms at which point the shock
begins to expand rapidly, signaling the onset of explosion.

The second panel of Fig.~\ref{fig:rshock} shows a measure of the
overall shock asymmetry, the normalized standard deviation of the
shock radius $\tilde{\sigma}$. 
The shock asymmetry grows as n5m2 \fheat 1.02
experiences runaway shock expansion, indicating that the explosion is
aspherical, as is also clear from the bottom-right panel of
Fig.~\ref{fig:volRend}.  The failed explosions show comparatively
small values of $\tilde{\sigma}$, implying relative sphericity of the
shock surface, until strong SASI oscillations set in after the shock
has receded \citep[see][]{Couch:2013leak}. 

The presence of pre-shock perturbations has substantial impact on the neutrino heating efficiency, $\eta = Q_{\rm net} (L_{\nu_e} + L_{\bar{\nu_e}})^{-1}$.
As shown in the third panel of Fig.~\ref{fig:rshock}, for n5m2 \fheat 1.00, the heating efficiency history is very similar to that of n0m0 \fheat 1.02.
This implies that the perturbations drive nonradial motion that increases the dwell time of material in the gain region, significantly enhancing the fraction of neutrino luminosity absorbed.
For n5m2 \fheat 1.02, the combination of \fheat $> 1$ {\it and} pre-shock perturbations results in a sufficiently increased heating efficiency to initiate a neutrino-driven explosion.
Also, $\eta$ depends sensitively, and non-linearly, on \fheat.
The time-averaged heating efficiencies for simulations n0m0 \fheat 1.00, n0m0 \fheat 1.02, n5m2 \fheat 1.00, and n5m2 \fheat 1.02 are 0.062, 0.080, 0.075, and 0.100, respectively.

It is almost exactly at the positive inflection in the average shock
radius curve of n5m2 \fheat 1.02 ($\sim 200$ ms) that the critical
condition for explosion, $\tau_{\rm adv} / \tau_{\rm heat} > 1$ is
satisfied (Fig.~\ref{fig:rshock}; \citealt{Thompson:2000gd,
  {Janka:2001fp}, {Buras:2006hl}, {Fernandez:2012kg}}).  Here we
define the average advection time through the gain region as
$\tau_{\rm adv} = M_{\rm gain} / \dot{M}$ and the gain region heating
time as $\tau_{\rm heat} = \abs{E_{\rm gain}} / Q_{\rm net}$, where
$\abs{E_{\rm gain}}$ is the total specific energy of the gain region
and $Q_{\rm net}$ is the net neutrino heating in the gain region
\citep[c.f.][]{{Muller:2012gd}, {ott:13a}}.  During the stalled-shock
phase of n5m2 \fheat 1.02, around $100 - 200\,\mathrm{ms}$, the ratio
$\tau_{\rm adv} / \tau_{\rm heat}$ is growing continuously.  Once this
critical ratio exceeds unity, thermal energy builds up in
the gain region faster than it can be advected out into the cooling
layer and the shock begins to expand.

\begin{figure}[tb]
  \centering
  \includegraphics[width=3.4in,trim= 0in 0.in 0.2in 0.1in, clip] {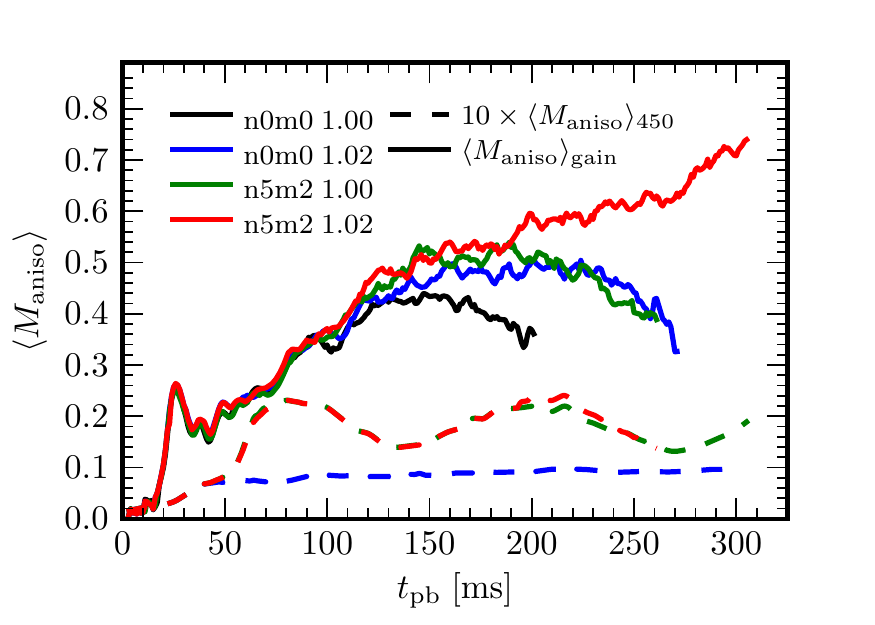}
  \caption{
    Density-weighted average of the Mach number of anisotropic motion [Eq.~(\ref{eq:maniso})] in two separate regions: the gain region (solid lines) and a 100-km wide spherical shell centered on $r=450$ km (dashed lines, multiplied by 10).
  }
  \label{fig:maniso}
\end{figure}

In order to assess the magnitude of the perturbations as they are
actually impinging upon the shock, and their effect on
the turbulent postbounce flow, we compute the density-weighted
radial averages of the Mach number of anisotropic motion,
\begin{equation}
  \langle M_{\rm aniso} \rangle = \left \langle \frac{v_{\rm aniso}}{\langle c_S \rangle_{4\pi}} \right \rangle_r,
  \label{eq:maniso}
\end{equation}
where the sound speed is first angle-averaged and the velocity of anisotropic motion has the same definition as in \citet{ott:13a, Couch:2013leak}.
The anisotropic Mach numbers for the gain layer and for
the preshock radial interval $400-500$ km are shown in
Fig.~\ref{fig:maniso}.  The differences between $\langle M_{\rm
  aniso} \rangle_{450}$ for perturbed and unperturbed cases are
evident.  The unperturbed cases, n0m0 \fheat 1.00 and n0m0 \fheat
1.02, have $\langle M_{\rm aniso} \rangle_{450} \lesssim 0.01$,
whereas in the perturbed models we find peak values of $\langle M_{\rm
  aniso} \rangle_{450}$ of $\gtrsim$0.02.  The Mach number of the
perturbations is dramatically reduced by the compression resulting
from infall toward the shock.  Larger pre-shock values of $\langle
M_{\rm aniso} \rangle$ correlate with larger post-shock values of
$\langle M_{\rm aniso} \rangle$.
The perturbed models for both low and high heat factors show similarly
large values of $\langle M_{\rm aniso} \rangle_{\rm gain}$ until
$\sim200$ ms when n5m2 \fheat 1.02 begins to explode.  The unperturbed
models have lower values of $\langle M_{\rm aniso} \rangle_{\rm gain}$
than either perturbed model.  The Mach number of anisotropic motion
for n0m0 \fheat 1.02 overtakes that of n5m2 \fheat 1.00 around 220 ms,
which we attribute to stronger neutrino-driven
convection.

\begin{figure}[htb]
  \centering
  \includegraphics[width=3.4in,trim= 0in 0.3in 0.1in 0.3in, clip] {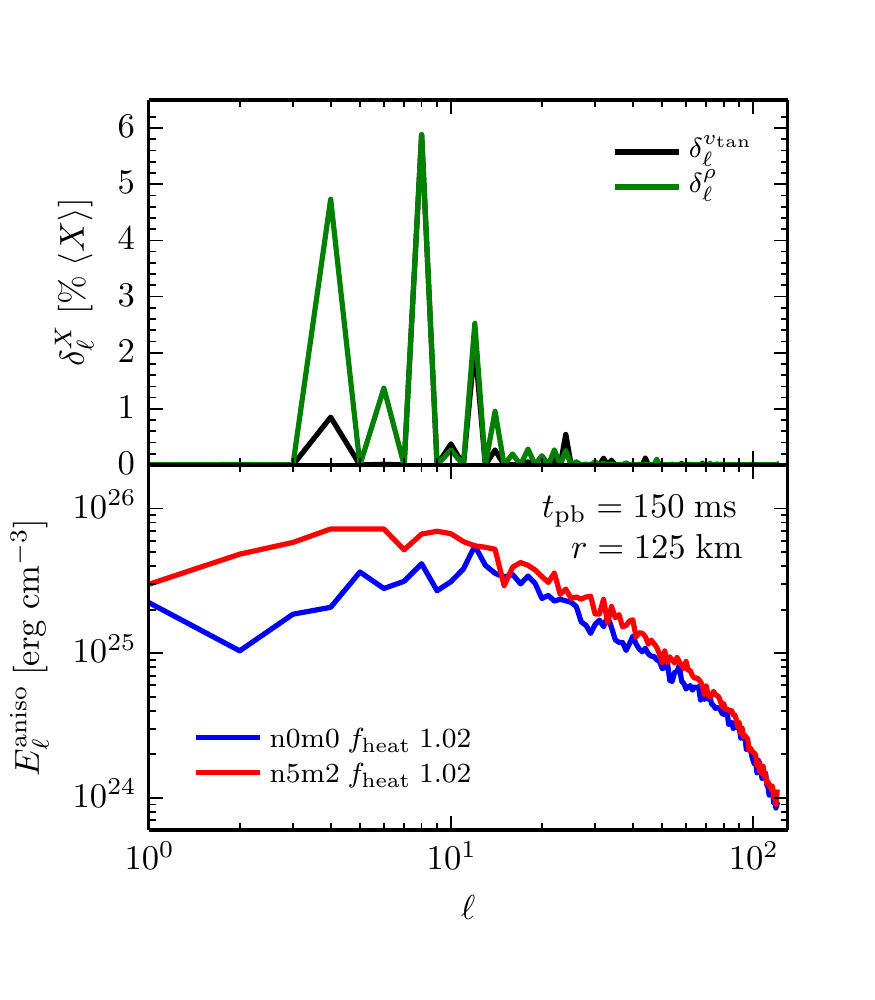}
  \caption{
    Power spectra in spherical harmonic basis of the perturbations in the pre-shock accretion flow (top).
    The perturbation spectra are computed as the difference of the sums of the squared spherical harmonic coefficients between the perturbed and unperturbed models, n5m2 \fheat 1.02 and n0m0 \fheat 1.00 [Eq.~(\ref{eq:pell})].
    The spectra are computed within a 10 km-wide shell centered on $r=400$ km and averaged over the 10 ms around $t_{\rm pb} = 100$ ms.
    Shown are the normalized perturbation spectra of the tangetial velocity (black lines), and of the density field (green lines).
    We normalize $\delta_\ell^{v_{\rm tan}}$ by the spherically-averaged {\it radial} velocity at $r=400$ km.
    The bottom panel shows the anisotropic kinetic energy spectra in the gain region at $t_{\rm pb} = 150$ ms.
    The spectra are averaged over a 10 km-wide shell centered on $r=125$ km and averaged over 10 ms.
    The anisotropic kinetic energy spectra give a measure of the effect of the perturbations on the nonradial flow in the gain region.
  }
  \label{fig:pertSpec}
\end{figure}

Another useful metric of the character of the disturbances reaching
the shock is the power spectrum of the perturbations, which we show in
Fig.~\ref{fig:pertSpec}.  We define the power spectrum of the
perturbations to a scalar field $X$ as $\delta_\ell^X = P_\ell^{X,{\rm perturbed}} -
P_\ell^{X,{\rm unperturbed}}$, where
\begin{equation}
  P_\ell^X = \sum_{m=-\ell}^\ell \left [ \oint X(\theta,\phi) Y_\ell^m (\theta,\phi) d\Omega \right ]^2.
\label{eq:pell}
\end{equation}
The spherical harmonics, $Y_\ell^m$, have their usual definition, and
details of similar calculations may be found in, e.g.,
\citet{{Hanke:2012dx}, {Dolence:2013iw}, {Couch:2013fh}}.  Figure
\ref{fig:pertSpec} shows the perturbations for tangential velocity,
$v_{\rm tan} = \sqrt{v_\theta^2 + v_\phi^2}$, and density, where we
have set $X$ to the square root of these quantities so $\delta_\ell^X$
has units of velocity and density, respectively.  We normalize
$\delta_\ell^{v_{\rm tan},\rho}$ by the angle-averaged radial velocity
and density at 400 km, respectively.  The applied perturbations to
$v_\theta$ manifest themselves in peak values of $\delta_\ell^{v_{\rm
    tan},\rho}$ of $\sim6\%$ at $\ell = 8$.  Nonradial motion caused
by the initial perturbations results in the growth of density
contrasts during infall \citep[][]{lai:00}.  The peak
  values of $\delta_\ell^{\rho}$ correspond to density contrasts
  reaching the shock front of $\gtrsim 2\times10^6\,\mathrm{g\, cm}^{-3}$.

Also shown in Fig.~\ref{fig:pertSpec} is a powerful diagnostic of the
strength of convective and turbulent motions in the gain region, the
spectrum of anisotropic kinetic energy, $E_\ell^{\rm aniso}$.  It is
computed from Eq.~(\ref{eq:pell}) with $X= \sqrt{\rho [(v_r -
    \langle v_r \rangle_{4\pi})^2 + v_\theta^2 + v_\phi^2 ]}$.  Model
n5m2 \fheat 1.02 has {\it significantly} more anisotropic kinetic
energy at large scales than the unperturbed simulation, n0m0 \fheat
1.02.  Above $\ell \approx 10$, the spectra of the perturbed and
unperturbed cases become fairly similar.  The more dramatic difference
at small $\ell's$ corresponds to the spatial scales of the
perturbations that are reaching the shock, as measured by
$\delta_\ell$.  Kinetic energy on large scales has been noted to
correlate with conditions favorable for explosion in a
number of previous studies \citep[e.g.,][]{{Hanke:2012dx},
  Couch:2013fh}.

In summary, the message of the various analyses we
  present in Figs.~\ref{fig:rshock}--\ref{fig:pertSpec} is clear:
  models with perturbations develop more vigorous postbounce
  turbulence, have higher neutrino heating efficiencies, and either
  explode or are much closer to explosion than their unperturbed
  counterparts. It is particularly noteworthy that the perturbations
  boost model n5m2 \fheat 1.00 to essentially the same heating
  efficiency and shock radius evolution as the unperturbed, more
  strongly heated model n0m0 \fheat 1.02. As pointed out by
  \cite{Foglizzo:2006js} and first demonstrated by
  \cite{Scheck:2008ja}, the development and strength of
  neutrino-driven convection in the gain layer increases with
  increasing magnitude of the accreting seed perturbations.  Stronger
  nonradial motion increases the dwell time of material in the gain
  layer. Thus, our models with perturbations absorb neutrino energy
  more efficiently, which favors explosion \citep[c.f.][]{{Thompson:2005iw}, {Murphy:2008ij}}.

\section{Conclusions}

The final phase of nuclear burning in massive stars approaching core
collapse is fast and furious. The Si/O shells surrounding the iron
core are sites of large-scale deviations of turbulent flow from
spherical symmetry. Our 3D postbounce CCSN simulations show that aspherical
perturbations in the Si/O layer can have important effects on the 3D
hydrodynamics of CCSNe. They lead to more vigorous turbulent flow behind
the shock and \emph{qualitatively} alter the outcome of core collapse:
they can turn a dud into an explosion.

The nonradial momentum-preserving velocity perturbations that we
considered here have spatial frequency and Mach numbers comparable to
what is expected from 2D Si/O burning simulations \citep{bazan:98,
  {Arnett:2011ga}}. These perturbations are mild compared to the large
$\ell=1$ density variations imposed by the previous studies of
\cite{bh:96} and \cite{fryer:04kick}. 

Our simulations \emph{prove the principle} that nonradial velocity
perturbations from convective Si/O burning can alter postbounce CCSN
hydrodynamics and can affect the explosion mechanism.
We study the effect of only one particular
perturbation, however, it is likely that the outcome will depend
on both magnitude and spatial dependence of the perturbations.
This must be explored in future work.
The 3D oxygen burning simulations of \cite{Meakin:2007dj} suggest that
in 3D the Mach numbers of fluctuations may be only half as large as in
2D. However, \cite{Meakin:2007dj} included only the O shell in 3D and
\citet{Arnett:2011ga} argue that it is the interplay of Si and O
burning shells that drives the most violent fluctuations. Thus, we
feel that our Mach 0.2 perturbations in the Si/O layer in 3D are {\it
  plausible}.

Recent studies comparing 2D and 3D CCSN hydrodynamics suggest that
explosions are more readily obtained in 2D than in 3D
(\citealt{Hanke:2012dx,{Couch:2013leak},{Couch:2013fh},
  {Hanke:2013kf},{Takiwaki:2013ui}}, but see \citealt{Dolence:2013iw}
for a differing view). CCSN theory, however, must robustly produce and
explain explosions in 3D to match observations. There are efforts
underway by many groups to improve upon current 3D simulations in
treatments of neutrino transport, weak interaction physics, magnetic
fields, and gravity with the hope of robustly producing explosions in
3D.  Our work shows that the initial conditions also matter, reminding
us that the CCSN mechanism is essentially an initial value problem.
At least part of the solution to the long-standing supernova problem
must lie in multi-D progenitor structure. Full-core, full-3D
progenitor evolution simulations to the onset of iron core collapse
are urgently needed.

\section*{Acknowledgements}
We acknowledge helpful discussions with E.~Abdikamalov, D.~Arnett,
P.~Goldreich, C.~Graziani, C.~Meakin, E.~O'Connor, U.~C.~T.~Gamma,
C.~Reisswig, L.~Roberts, and N.~Smith.  SMC is 
supported by NASA through Hubble Fellowship grant No.\ 51286.01 awarded
by the Space Telescope Science Institute.
CDO is partially supported by
NSF grant nos.\ AST-1212170, PHY-1151197, and OCI-0905046 and by
the Alfred~P.~Sloan Foundation.

The software used in this work was in part developed by the DOE
NNSA-ASC OASCR Flash Center at the University of Chicago.  The
simulations were carried out on computational resources at ALCF at
ANL, which is supported by the Office of Science of the US Department
of Energy under Contract No. DE-AC02-06CH11357, and on the NSF
XSEDE network under computer time allocation TG-PHY100033.


\end{document}